\documentclass[12pt]{revtex4}
\usepackage{amsmath}
\usepackage{amssymb}
\usepackage{epsfig}

\begin{document}

\title{Magnetic properties of antiferromagnetic quantum Heisenberg 
spin systems with a \emph{strict} single particle site occupation.} 
\author{Raoul Dillenschneider}
\email{rdillen@lpt1.u-strasbg.fr}
\author{Jean Richert}
\email{richert@lpt1.u-strasbg.fr}
\affiliation{
Laboratoire de Physique Th\'eorique, UMR 7085 CNRS/ULP,\\
67084 Strasbourg Cedex,France}

\begin{abstract}
We work out the magnetization and susceptibility of Heisenberg- and
 XXZ-model antiferromagnet spin-1/2 systems in $D$ dimensions 
under a rigorous constraint of single particle site occupancy. Quantum 
fluctuations are taken into account up to the first order in a loop expansion 
beyond the N\'eel state mean field solution. We discuss the results, their 
validity in the vicinity of the critical point and compare them with the 
results obtained by means of a spin wave approach.
\end{abstract}

\maketitle
\section{Introduction}

Recent work on quantum spin systems discuss the possible existence of
spin liquid states and in two-dimensional space dimensions
the competition or phase transition between spin liquid states and an 
antiferromagnetic N\'eel state which is naturally expected to describe 
Heisenberg type systems \cite{SenthilFisher-04,SenthilFisher-05,
GhaemiSenthil-05,Morinari-05}. It is also known that undoped superconducting 
systems show an antiferromagnetic phase \cite{LeeNagaosaWen-04}.

In the following we focus our attention on a N\'eel phase description of 
quantum spin systems described by Heisenberg models. 
More precisely we present below a detailed study of the magnetization and the 
parallel magnetic susceptibility of Heisenberg antiferromagnetic spin-1/2 
systems on $D$-dimensional lattices at finite temperature. The aim of the 
work is the study of the physical pertinence of the N\'eel state ansatz as a 
mean-field approximation in the temperature interval $0<T<T_c$ where $T_c$  is 
the critical temperature. In order to get a precise answer to this point we work
out the quantum fluctuation contributions beyond the mean-field approximation 
under the constraint of \emph{strict} single site occupancy \cite{Popov-88} 
which allows to avoid a Lagrange multiplier approximation \cite{Auerbach-88}. 
The results are also extended to anisotropic $XXZ$ systems and compared to those
obtained in the framework of the spin wave approach.

The paper is organized as follows.
In section II we present the derivation of the partition function under the
single particle site occupation constraint. The mean-field and first order
loop expansion term contributions are derived in section III. In section IV we 
determine the magnetization and the magnetic susceptibility, and discuss the 
results obtained at the different levels of approximation. Comments are 
presented and conclusions are drawn in section V. Details of calculations are 
presented in the appendix, section VI. 

\section{Fermionization of the Heisenberg model and the partition function}

The Heisenberg antiferromagnet Hamiltonian (HAFM) in the presence of a local
magnetic field $\vec{B}_i$ reads

\begin{equation}
H = -\frac{1}{2} \underset{<i,j>}{\sum} 
J_{ij} \vec{S}_i.\vec{S}_j + \underset{i}{\sum}
\vec{B}_i .\vec{S}_i
\end{equation}
where ${J_{ij}<0}$ and the sums in the first term run over nearest-neighbour 
sites $<i,j>$ on a $D$-dimensional hypercubic lattice.  
 
The $S=1/2$ spin vector operators are expressed in terms of fermionic creation 
and annihilation operators $\{f_{i\lambda}, f^{\dagger}_{i\alpha}\}$ 

\begin{eqnarray}
\vec{S}_i = f^{\dagger}_{i\alpha} \vec{\sigma}_{\alpha\lambda} f_{i\lambda}
\end{eqnarray}
where the $\vec{\sigma}_{\alpha \lambda}$ vector components are Pauli matrices.

The transformation is rigorous if $\sum_\alpha f^{\dagger}_{i\alpha}
f_{i\alpha}=1$. The Fock space constructed with the fermionic operators 
$f,f^{\dagger}$ is not in bijective correspondence with the Hilbert space of 
the spin states. Indeed, in Fock space and for spin-1/2 particles, the 
occupation of each site $i$ can be characterized by the states 
$|n_{i,\uparrow},n_{i,\downarrow}>$ with $n_{i,\alpha} \in \{0,1\}$, that is 
states $|0,0>$, $|1,0>$, $|0,1>$ and $|1,1>$. But in the case of single
occupancy the states $|0,0>$ and $|1,1>$ which are excluded as \emph{unphysical}
in the present case have to be eliminated. This is done by means of a projection
procedure proposed by  Popov and Fedotov \cite{Popov-88} and generalized to 
$SU(N)$ symmetry in ref. \cite{KFO-01}.

Introducing the projection operator $\tilde{P}= e^{i \frac{\pi}{2} \tilde{N}}$
where $\tilde{N}=\underset{i,\sigma}{\sum}f^{\dagger}_{i\sigma} f_{i\sigma}$
is the number operator the partition function $\cal{Z}$ reads

\begin{equation*}
\mathcal{Z}=Tr \left[ e^{-\beta H}\tilde{P} \right]
\end{equation*}

where $\beta$ is the inverse temperature. On each site $i$ the contributions 
of states $|0,0>$ and $|1,1>$ to  $\cal{Z}$ eliminate each other. Indeed 
 
\begin{gather*}
<0,0|_i e^{-\beta H}.e^{i\frac{\pi}{2}*0}  |0,0>_i +
<1,1|_i e^{-\beta H}.e^{i\frac{\pi}{2}*2} |1,1>_i \\
+ <1,0|_i e^{-\beta H}.e^{i\frac{\pi}{2}} |1,0>_i
+ <0,1|_i e^{-\beta H}.e^{i\frac{\pi}{2}} |0,1>_i \\
= i(<1,0|_i e^{-\beta H} |1,0>_i
+ <1,0|_i e^{-\beta H} |1,0>_i )
\end{gather*} 
 
Hence the partition function  
 
\begin{eqnarray}
\mathcal{Z}=\frac{1}{i}.
Tr \left[ e^{-\beta (H-\mu\tilde{N})} \right]
\end{eqnarray} 
with the imaginary "chemical potential" $\mu = i \frac{\pi}{2 \beta}$ 
describes a system with strictly one particle per lattice site, in contrast 
with the usual method which introduces an average projection by means of 
a real Lagrange multiplier \cite{Auerbach-88,Auerbach-94}.

\section{Mean field and one-loop approximations}

Following the usual procedure we transform the Heisenberg Hamiltonian 
into a bilinear fermionic expression using a Hubbard-Stratonovich
decoupling. Starting from (3) this leads to 

\begin{equation}
\mathcal{Z}= \frac{1}{\mathcal{Z}_0}
\int \prod_i \mathcal{D}  \vec\varphi_i
\int_{\xi_{i\sigma}(\beta)=i\xi_{i\sigma}(0)} 
\mathcal{D} (\xi_{i\sigma}^{*},\xi_{i\sigma})
e^{-\int_0^\beta d\tau \left[
 \underset{i,\sigma}{\sum}
\xi_{i\sigma}^{*} \frac{\partial}{\partial \tau} \xi_{i\sigma} 
+ S_0 \left[ \varphi(\tau) \right] +
\sum_i \vec{\varphi}_i.\vec{S}_i(\tau) \right] }
\label{Zgrassmann}
\end{equation}
where $\tau$ is an imaginary time and

\begin{eqnarray*}
\mathcal{Z}_0
&=& \int \prod_i \mathcal{D} \vec\varphi_i 
e^{-\int_0^\beta d\tau S_0 \left[ \vec\varphi(\tau) \right]} \\
S_0 \left[ \vec\varphi(\tau)  \right] &=&
\frac{1}{2} \sum_{i,j} 
J_{ij}^{-1} 
(\vec{\varphi}_i(\tau)-\vec{B}_i).(\vec{\varphi}_j(\tau)-\vec{B}_j)
\end{eqnarray*}
where $\vec{\varphi}$ stands for the Hubbard-Stratonovich decoupling fields and 
$\xi$  for the \emph{Grassmann} variables.

After integration over the bilinear fermionic $\{\xi_{i  \sigma}\}$
terms which appear in the action $\cal{Z}$ takes the form

\begin{equation*}
\mathcal{Z} 
= \frac{1}{Z_0} \int \mathcal{D} \vec\varphi e^{-S_{eff}\left[ \vec\varphi 
\right]}
\end{equation*}

where 

\begin{eqnarray}
S_{eff}\left[ \vec\varphi \right] 
= \int_0^\beta d\tau S_0 \left[ \vec\varphi(\tau) \right]
-\sum_i ln \, 2 ch \frac{\beta}{2} \| \vec{\bar{\varphi}}_i(\omega=0) \|
+ Tr\{ \sum_{n=1}^{\infty} \frac{1}{n} (G_0 M_1)^n \}
\label{Seff}
\end{eqnarray}
and $(\vec{\bar{\varphi}}_i(\omega)=0)$ is the Fourier transform of 
$\vec{\bar{\varphi}}_i(\tau)$. The propagator $G_0$ and $M_1$ are defined 
in matrix form in appendix \ref{G0M1}.

In a loop expansion beyond the mean-field approximation $\vec{\bar{\varphi}}$ 
the effective action given by (\ref{Seff}) is expanded in a Taylor
series
 
\begin{eqnarray*}
S_{eff}\left[ \vec{\varphi} \right] =
{S_{eff}} {\Big\vert}_{\left[\vec{\bar{\varphi}}\right]}
+ \frac{\partial S_{eff}} {\partial \vec{\varphi}}
 {\Big\vert}_{\left[\vec{\bar{\varphi}}\right]} \delta \vec{\varphi}
+ \frac{1}{2} \frac{\partial^2 S_{eff}} {\partial \vec{\varphi}^2}
{\Big\vert}_{\left[\bar{\vec{\varphi}}\right]}
 {\delta \vec{\varphi}}^2 + \mathcal{O} (\delta \vec{\varphi}^3)
\end{eqnarray*}
to second order (one-loop contribution) in the fluctuations 
$\vec{\delta \varphi}^2$
of $\vec{\varphi} = \vec{\bar{\varphi}} + \delta \vec{\varphi}$. 
Since the mean 
field $\vec{\bar{\varphi}}$ is chosen in such a way that 
$\frac{\partial S_{eff}}{\partial \vec{\varphi}}
{\Big\vert}_{\left[\bar{\vec{\varphi}}\right]} \delta \vec{\varphi} = 0$ one 
gets the set of coupled self-consistent equations 

\begin{eqnarray*}
\sum_j J_{ij}^{-1} 
\left[\vec{\bar{\varphi}}_j-\vec{B}_j \right]
=\frac{1}{2} \frac{\vec{\bar{\varphi}}_i}{\bar{\varphi}_i}
th \left[ \frac{\beta\bar{\varphi}_i}{2} \right]
\end{eqnarray*}
which fixes the fields $\vec{\bar{\varphi}}$.

In the following we consider a N\'eel mean-field order
$\vec{\bar{\varphi}}_i(\tau) = (-1)^{\vec{\pi}.\vec{r}_{i}} 
\bar{\varphi}^z \vec{e}_z = \bar{\varphi}^z_i \vec{e}_z$ where 
$\vec{\pi}$ is the Brioullin spin sublattice vector.
The magnetic field applied to the system is also chosen to be aligned along 
the direction $\vec{e}_z$. The partition function can be decomposed into a 
product of three terms

\begin{eqnarray*}
\mathcal{Z} = {\mathcal{Z}_{MF}}.{\mathcal{Z}_{zz}}
.{\mathcal{Z}_{+-}}
\end{eqnarray*}

where $\mathcal{Z}_{MF}$, $\mathcal{Z}_{zz}$ and
$\mathcal{Z}_{+-}$ are given by

\begin{eqnarray*}
\mathcal{Z}_{MF} &=& e^{-{S_{eff}} {\Big\vert}_{\left[\bar
{\varphi}\right]} }
 \\
\mathcal{Z}_{zz} &=& \frac{1}{\mathcal{Z}^{zz}_{0}}
\int \mathcal{D}\varphi^z e^{-
\frac{1}{2} \frac{\partial^2 S_{eff}}{{\partial \varphi^z}^2}
{\Big\vert}_{\left[\bar{\varphi}\right]}
 {\delta \varphi^z}^2 }
 \\
\mathcal{Z}_{+-} &=& \frac{1}{\mathcal{Z}_0^{+-}}
\int \mathcal{D}(\varphi^{+},\varphi^{-}) e^{-
\frac{1}{2} \frac{\partial^2 S_{eff}}{{\partial \varphi^{+}
\partial \varphi^{-}}}
{\Big\vert}_{\left[\bar{\varphi}\right]}
 {\delta \varphi^{+}}.{\delta \varphi^{-}} }
\end{eqnarray*}

with 

\begin{equation*}
{S_{eff}} {\Big\vert}_{\left[\bar{\varphi}\right]}
 =
\frac{\beta}{2} \sum_{i,j} J_{ij}^{-1} 
\left[(\bar{\varphi}^z_i-B^z_i)
.(\bar{\varphi}^z_j-B^z_j) \right]
-\sum_i ln \, 2 ch \frac{\beta}{2} 
\| \bar{\varphi}^z_i \|
\end{equation*}

\begin{align}
\frac{1}{2} \frac{\partial^2 S_{eff}}{{\partial \varphi^z}^2}
{\Big\vert}_{\left[\bar{\varphi}\right]}
 {\delta \varphi^z}^2
= &
\sum_\omega \sum_{i,j} \frac{\beta}{2}
\left[
 J_{ij}^{-1} - 
\left(
\frac{\beta}{4} 
th^{'} \left( \frac{\beta}{2} \bar{\varphi}_i^z \right)\right)
\delta_{ij} \delta(\omega=0) \right]
\delta \varphi_i^z(-\omega)\delta \varphi_j^z(\omega)
\notag \\
\frac{1}{2} \frac{\partial^2 S_{eff}}{{\partial \varphi^{+}
\partial \varphi^{-}}}
{\Big\vert}_{\left[\bar{\varphi}\right]}
 {\delta \varphi^{+}} {\delta \varphi^{-}}
= &
\sum_\omega \sum_{i,j} \frac{\beta}{2}
\left[
\frac{1}{2} J_{ij}^{-1} - 
\left( \frac{1}{2} 
\frac{th \left( \frac{\beta}{2} \bar{\varphi}_i^z \right) }
{\bar{\varphi}_i^z - i \omega } \right)
\delta_{ij} \right]
\delta \varphi_i^{+}(-\omega)\delta \varphi_j^{-}(\omega)
\notag \\
+ &
\sum_\omega \sum_{i,j} \frac{\beta}{2}
\left[ \frac{1}{2} J_{ij}^{-1} \right]
\delta \varphi_i^{+}(\omega) \delta \varphi_j^{-}(-\omega)
\label{delta2Seff}
\end{align}
$\mathcal{Z}_{MF}$ is the mean field contribution, $\mathcal{Z}_{zz}$
and $\mathcal{Z}_{+-}$ are the one-loop contributions 
respectively for the longitudinal part $\delta \varphi^z$ and
the transverse parts of $\vec{\varphi}$, $\delta \varphi^{+-}$,
which take account of the fluctuations around the mean-field value 
$\bar{\varphi}^z$. 
 
The contributions $\mathcal{Z}_{zz}$ and $\mathcal{Z}_{+-}$ 
are quadratic in the 
field variables $\delta \varphi^z, \delta \varphi^{+-}$ and can
be worked out in the presence of a staggered magnetic field $B_i^z$. 
Studies involving a uniform magnetic field acting on 
antiferromagnet quantum spin systems can also be found in ref.\cite{KFO-01}.

\section{Magnetization and susceptibility of d-dimensional systems}

\subsection{Magnetization}

The fields $\{\vec{\bar{\varphi}}_i\}$ can be related to the magnetizations
$\{\vec{\bar{m}}_i\}$ as shown in appendix \ref{ApB} and the free energy can be 
expressed in terms of this order parameter, see appendix \ref{ApC}. 
The magnetization $m$ on site $i$ is 
the sum of a mean field contribution $\bar{m} = -\frac{1}{\beta} 
\frac{\partial ln \mathcal{Z}_{MF}}{\partial B^z}$, a transverse contribution 
$\delta m_{+-} = -\frac{1}{\beta} \frac{\partial ln \mathcal{Z}_{+-}}
{\partial B^z}$ and a longitudinal contribution $\delta m_{zz} = -\frac{1}
{\beta} \frac{\partial ln \mathcal{Z}_{zz}}{\partial B^z}$. For a small magnetic
field $\vec B$ a linear approximation leads to 
$m = \bar{m} + \delta m_{zz} + \delta m_{+-}$ where
\begin{eqnarray*}
\bar{m} &=& \frac{1}{2} th \frac{\beta}{2}D|J|\bar{m}
 \\
\delta m_{zz} &=&
 -\frac{1}{N_p \beta} \sum_{\vec{k} \in SBZ}
\frac{8\bar{m} \Delta \tilde{m}_0 \left(1-4\bar{m}^2\right)
\left(\frac{\beta D |J| \gamma_{\vec{k}}}{2}\right)^2}
{\left[1 -\left(\frac{\beta D |J| \gamma_{\vec{k}}}{2}\right)^2 
 \left(1-4\bar{m}^2\right)^2\right]}
 \\
\delta m_{+-} &=&
\frac{\left(1 + 2 D |J| \Delta \tilde{m}_0 \right)}{4\bar{m}}
- \frac{1}{N_p} \sum_{\vec{k} \in SBZ}
\frac{\left(1 + 2 D |J| \Delta \tilde{m}_0 (1 - \gamma_{\vec{k}}^2)\right)}
{\sqrt{1 - \gamma_{\vec{k}}^2}}
\frac{1}{\left[\text{th} \left(\beta D |J| \bar{m}
\sqrt{1 - \gamma_{\vec{k}}^2} \right) \right]}
\end{eqnarray*}
$N_p$ is the number of spin-1/2 sites, 
$\Delta \tilde{m}_{0} = \frac{\frac{\beta}{4}\left(1-4.\bar{m}^2 \right)}
{1-\frac{\beta}{2}D|J|\left(1-4.\bar{m}^2 \right)}$ and 
$\gamma_{\vec{k}} = \frac{1}{D} \underset{\vec{\eta} \in n.n.}{\sum} 
\cos (\vec{k}.\vec{\eta}) $, see appendix \ref{ApC} for details of the
derivation.

At low temperature $(T \rightarrow 0)$ the magnetization goes over to the 
corresponding spin-wave expression 
\cite{Igarashi-92,Manousakis-91,Holstein-40,Azakov-01}, which reads

\begin{eqnarray*}
m = 1 - \frac{1}{N_p} \sum_{\vec{k} \in SBZ} 
\frac{1}{\text{th } 
\left( \frac{\beta D |J|}{2}\sqrt{1-\gamma_{\vec{k}}^2} \right) }
.\frac{1}{\sqrt{1-\gamma_{\vec{k}}^2}}
\end{eqnarray*}

\begin{figure}
\centering
\mbox{\epsfig{file=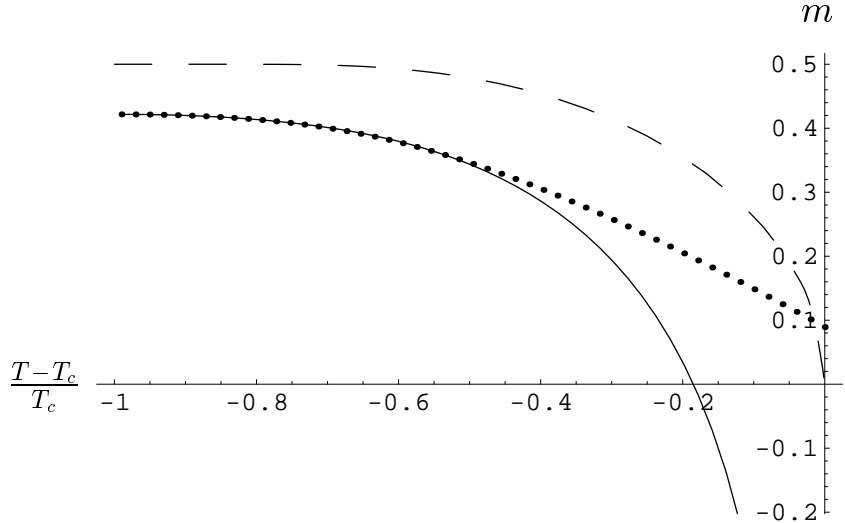}}
\caption{Magnetization in a 3D Heisenberg antiferromagnet cubic 
lattice. 
Dashed line : Mean field magnetization, 
Dotted line : Spin wave magnetization, 
Full line : One-loop corrected magnetization.}
\label{fig:MagP3D100}
\end{figure}

Figure \ref{fig:MagP3D100} shows the magnetization $m$ in the mean-field,
the one-loop and the spin wave approach for temperatures $T \leq T_c$
where $T_c = D|J|/2$ corresponds to the critical point. One observes a sizable
contribution of the quantum fluctuations generated by the loop contribution 
over the whole range of temperatures as well as an excellent and expected 
agreement between the quantum corrected and the spin wave result at very low
temperatures. 
 
The magnetization shows a singularity in the neighbourhood of the critical
point. This behaviour can be read from the analytical expressions of 
$\delta m_{+-}$ and $\delta m_{zz}$ and is generated by the $|\vec{k}| = 0$  
mode which leads to $\gamma_{\vec{k}} = 1$ and by cancellation of $\bar{m}$. 
The N\'eel state mean-field 
approximation is a realistic description at very low $T$. With increasing
temperature this is no longer the case. The chosen ansatz breaks a symmetry
whose effect is amplified as the temperature increases and leads to the
well-known divergence disease observed close to $T_c$. Hence if higher order
contributions in the loop expansion cannot cure the singularity the N\'eel 
state antiferromagnetic ansatz does not describe the physical symmetries of the
system at the mean-field level at temperatures in the neighbourhood of the
critical point. Consequently it is not a pertinent mean-field approximation for
the description of the system.

\begin{figure}
\centering
\mbox{\epsfig{file=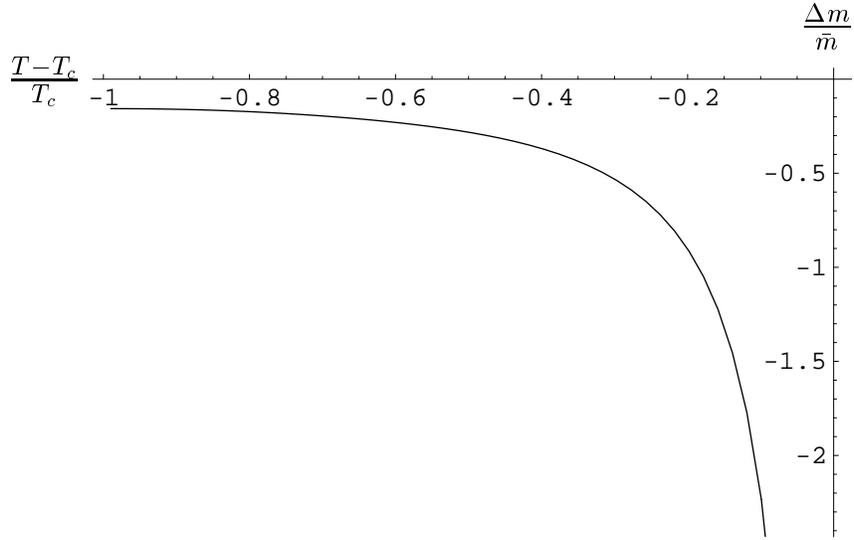}}
\caption{Ginzburg criterion $\frac{\Delta m}{\bar{m}}$ 
for the 3D Heisenberg model.}
\label{fig:GinzP3D100}
\end{figure}

The discrepancy can be quantified by means of the quantity $\frac{|\Delta m|}
{\bar{m}}$ where $\Delta m = m - \bar{m}=\delta m_{zz} + \delta m_{+-}$. 
Figure 2 shows the result. The 
relation $\frac{|\Delta m|}{\bar{m}} < 1$ (Ginzburg criterion) fixes a limit 
temperature $T_{lim}$ above which the quantum fluctuations generate larger 
contributions than the mean-field. For 3$D$ systems this leads to 
$T_{lim} \simeq 0.8 T_c$, see figure \ref{fig:GinzP3D100}.  

\begin{figure}
\centering
\mbox{\epsfig{file=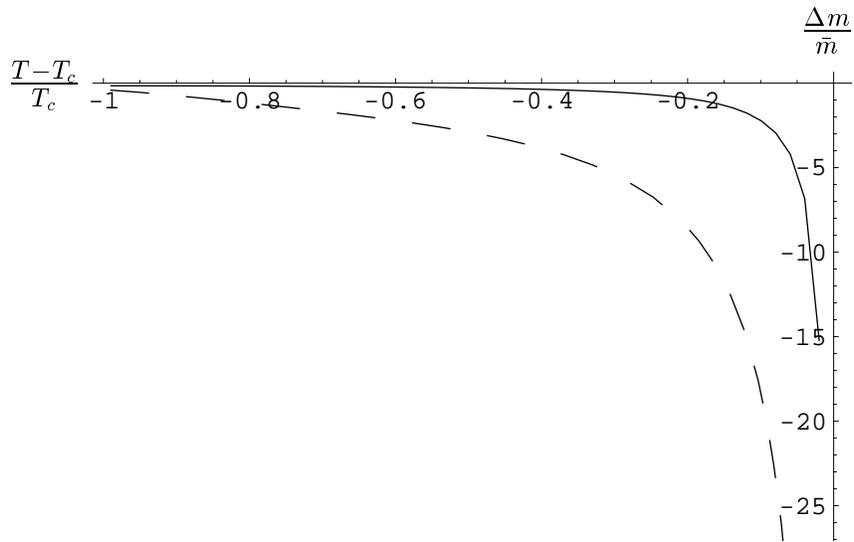}}
\caption{Comparison of the Ginzburg criterion
for 2D (dashed line) and 3D (full line) Heisenberg model.
$|\frac{\Delta m}{\bar{m}}|>1$ for $T \gtrsim 0$ at 2D
and $T \gtrsim 0.8 T_c$ at 3D.}
\label{fig:GinzDim}
\end{figure}

The pathology is the stronger the smaller the space dimensionality.
It is also easy to see on the expression of the magnetization that, 
as expected,
the contributions of the quantum fluctuations decrease with increasing $D$. 
As can be seen in the figure \ref{fig:GinzDim}, the saddle point breaks down
earlier in two than in three dimensions.
  
In fact,the Heisenberg model spin wave spectrum shows a Goldstone mode
as a consequence of the symmetry breaking by the N\'eel state. When   
$|\vec{k}|$ goes to zero 

\begin{eqnarray*}
\omega_{\vec{k}} &=& Z D S \sqrt{1 - \gamma_{\vec{k}}^2 } \\
\underset{\vec{k} \rightarrow \vec{0}}{\text{lim }}\omega_{\vec{k}} &\sim& 
 |\vec{k}|
\end{eqnarray*}
The zero mode destroys the long range order in 1D and 2D as expected from
the Mermin-Wagner theorem \cite{Mermin-66}.

In the case of the  
XXZ-model the Hamiltonian of the system can be written

\begin{eqnarray*}
H^{XXZ} = -\frac{J}{2} \sum_{<ij>} \left( S_i^x S_j^x+ S_i^y S_j^y 
+ (1+\delta)S_i^z S_j^z \right)
\end{eqnarray*}
where $\delta$ governs the anisotropy. In this case the excitation
spectrum shows a finite $|\vec{k}|=0$ energy $\omega_{\vec{k}}$

\begin{eqnarray*}
\omega_{\vec{k}} &=& Z D S \sqrt{1 - 
\left(\frac{J}{J+\Delta}\gamma_{\vec{k}}\right)^2 } \\
\underset{\vec{k} \rightarrow \vec{0}}{\text{lim }}\omega_{\vec{k}} &\sim& 
\sqrt{1 - \left(\frac{1}{1+\delta}\right)^2 \left(1 - \frac{\vec{k}^2}{2 D}
\right)}
\end{eqnarray*}

In appendix \ref{ApD} we develop explicitly the expressions of the free energy, 
magnetization and susceptibility. By examination  the expressions show that the
zero momentum mode is no longer responsible for a breakdown of the saddle point
procedure near $T_c^{XXZ} = \frac{D |J+\Delta|}{2}$ However the magnetization 
of the XXZ-model remains infinite near $T_c^{XXZ}$. This is due to
the common disease shared with the Heisenberg model that the mean field 
magnetization appearing in the denominator of $\delta m_{+-}$ goes to zero near
the critical temperature. One concludes that the mean-field N\'eel state 
solution makes only sense at low temperatures, that is  for 
$T \lesssim T_{lim}$, whatever the degree of symmetry breaking induced by the 
mean-field ansatz.

\subsection{Susceptibility}

We consider the parallel susceptibility $\chi_{\parallel}$ which characterizes 
a magnetic system on which a magnetic field is applied in the $Oz$ direction. 
The expression of $\chi_{\parallel}$ decomposes again into three contributions

\begin{eqnarray*}
\chi_\parallel = -\frac{1}{N_p} \frac{\partial^2 \mathcal{F}}{\partial B^2} 
\Bigg{\vert}_{B=0} =\chi_{MF} + \chi_{zz} + \chi_{+-} 
\end{eqnarray*}

with

\begin{align*}
{\chi_\parallel}_{MF} 
=& \Delta m_{\chi 0} = \frac{ \frac{\beta}{4}\left( 1 - 4\bar{m}^2 
\right) }{1+\frac{\beta}{2}D|J|\left(1-4\bar{m}^2 \right)}
 \\
{\chi_\parallel}_{zz}
=& -\frac{1}{N_p \beta} \sum_{\vec{k} \in SBZ}
\frac{8\left( \frac{\beta D |J| \gamma_{\vec{k}} }{2} \right)^2
\Delta m_{\chi 0}^2 \left(1 + 4 \bar{m}^2 \right)  }
{\left[ 1 - \left( \frac{\beta D |J| \gamma_{\vec{k}} }{2} \right)^2 
\left( 1 - 4 \bar{m}^2 \right)^2 \right]}
 \\
{\chi_\parallel}_{+-}
=&
 \frac{1}{N_p} \sum_{\vec{k} \in SBZ} \Bigg{\{}
- \frac{1}{2} \frac{\beta \left(1 - 2 D |J| \Delta m_{\chi 0} \right)^2}
{\text{sh }^2 \left(\beta D |J| \bar{m} \right)}
 \\
& 
+ \frac{1}{\text{sh }^2 \left(\beta D |J| \bar{m}\sqrt{1-\gamma_{\vec{k}}^2} 
\right)} 
 \\
& \Bigg{[}
\frac{\beta}{2} \left(1 - 2 D |J| \Delta m_{\chi 0} \right)^2
- \beta \left(D |J| \Delta m_{\chi 0} \gamma_{\vec{k}} \right)^2 
\frac{\text{sh }2\beta D |J| \bar{m}\sqrt{1-\gamma_{\vec{k}}^2} }
{\beta D |J| \bar{m}\sqrt{1-\gamma_{\vec{k}}^2} }
\Bigg{]}
\Bigg{\}}
\end{align*}

\begin{figure}
\centering
\epsfig{file=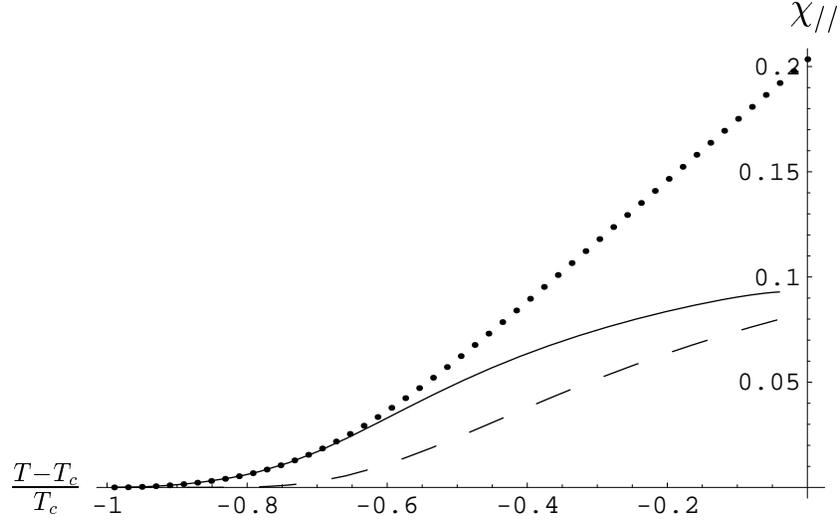}
\caption{Parallel magnetic susceptibility at 3D for Heisenberg model.
 Dotted line : Spin wave susceptibility. 
Dashed line : mean field susceptibility ${\chi_\parallel}_{MF}$.
Full line : total susceptibility (${\chi_\parallel}_{MF}
+\delta{\chi_\parallel}$).}
\label{Susceptibility}
\end{figure}

The behaviour of $\chi_{\parallel}$ is shown in Figure \ref{Susceptibility}
in which we compare the mean-field, spin wave and the one-loop corrected
contributions for a system on a 3 $D$ cubic lattice. 
One observes again a good agreement between the quantum corrected and the spin 
wave expressions at low temperatures. For higher temperatures the curves depart
as expected. The mean-field  contribution remains in qualitative agreement with 
the total contribution.
 
\section{Conclusion}

In the present work we showed the contribution of quantum fluctuations in the
description of quantum spin systems at finite temperature governed by 
Heisenberg Hamiltonians in space dimension $D$. The mean-field ansatz is taken 
as a N\'eel state. The number of particles per site is fixed by means of a 
rigorous constraint implemented in the partition function. It has been shown 
elsewhere \cite{Dillen-05} that this fact introduces a large shift of the 
critical temperature compared to the case where the constraint in generated 
through an ordinary Lagrange multiplier term. 

At low temperature the magnetization and the magnetic susceptibility are close 
to the spin wave value as expected, also in agreement with former work 
\cite{Azakov-01}. The quantum corrections are sizable even for low temperatures.
They increase with increasing temperature.    

At higher temperature the quantum contributions grow to a singularity in the
neighbourhood of the critical temperature. The assumption that the 
N\'eel mean-field contributes for a major part to the magnetization and the
susceptibility is no longer valid. 
Approaching $T_c$ the mean field contribution to the magnetization goes to 
zero and strong diverging fluctuations are generated at the one-loop order.
This behaviour is common to the Heisenberg and XXZ magnetization. In addition 
the N\'eel order breaks $SU(2)$ symmetry of the Heisenberg Hamiltonian inducing
low momentum fluctuations near $T_c$ which is not the case in the XXZ-model.

The influence of quantum fluctuations decreases with the dimension $D$ 
of the system due to the expected fact that the mean-field contribution 
increases relatively to the one-loop contribution. 

For dimension $D=2$ the magnetization verifies the Mermin and Wagner theorem 
\cite{Mermin-66} for $T \neq 0$, the fluctuations are larger than the mean field
contribution for any temperature. Therefore another mean-field ansatz has to be
found in order to describe the correct physics at finite and not too low 
temperature.\\

The authors would like to thank Drs. D.Cabra and T. Vekua for instructive 
discussions.

\section{Appendix}

\subsection{Matrices $G_0$ and $M_1$ \label{G0M1}}

After integration over the fermionic degrees of freedom in equation 
(\ref{Zgrassmann}) the partition function takes the form 

\begin{eqnarray*}
\mathcal{Z} &=& \frac{1}{Z_0} \int \mathcal{D} \varphi
e^{-\{ \int_0^\beta d\tau S_0 \left[ \varphi(\tau) \right] 
- ln \, det \left[ \beta M \right]  \}  } \\
&=& \frac{1}{Z_0} \int \mathcal{D} \varphi e^{-S_{eff}\left[ \varphi \right]}
\end{eqnarray*}

where
\begin{eqnarray*}
M_{i,(p,q)}=
\begin{bmatrix}
i p \delta_{p,q} + \frac{1}{2} \varphi_i^z(p-q) &
\frac{1}{2} \varphi_i^{-}(p-q) \\
\frac{1}{2} \varphi_i^{+}(p-q) &
i p \delta_{p,q} -\frac{1}{2} \varphi_i^z(p-q)
\end{bmatrix}
\end{eqnarray*}
$p$ and $q$ refer to modified fermionic Matsubara frequencies, 
$p=\omega_f - \mu =\frac{2\pi}{\beta}(n+1/4)$ and $n$ is an integer,
see \cite{Popov-88}.
$M$ can be put in the form

\begin{eqnarray*}
M &=& -G_0^{-1}(1-G_0 M_1)
\end{eqnarray*}

where
\begin{eqnarray*}
G_0 = 
\begin{bmatrix}
-\frac{1}{det\,G_p}
 \left[ i p \delta_{p,q} - \frac{1}{2} \bar{\varphi}_i^z(p-q=0)
\delta_{p,q} \right]
 &
\frac{1}{det\,G_p} \frac{1}{2} \bar{\varphi}_i^{-}(p-q=0)\delta_{p,q}
 \\
\frac{1}{det\,G_p} \frac{1}{2} \bar{\varphi}_i^{+}(p-q=0)\delta_{p,q}
 &
-\frac{1}{det\,G_p} \left[ i p \delta_{p,q} + \frac{1}{2} 
\bar{\varphi}_i^z(p-q=0)\delta_{p,q} \right]
\end{bmatrix}
\end{eqnarray*}

\begin{eqnarray*}
M_1 = 
\begin{bmatrix}
\frac{1}{2} \delta \varphi_i^z(p-q)
 &
\frac{1}{2} \delta \varphi_i^{-}(p-q)
 \\
\frac{1}{2} \delta \varphi_i^{+}(p-q)
 &
-\frac{1}{2} \delta \varphi_i^z(p-q)
\end{bmatrix}
\end{eqnarray*}
with $\delta \vec{\varphi}_i(p-q)=\vec{\varphi}_i(p-q)- 
\vec{\bar{\varphi}}_i(p-q=0)\delta_{p,q}$. The second term in the expression 
of $M$ corresponds to the quantum contributions.
The expression $ln \, det \, (\beta M)$ can be developed into a series  

\begin{eqnarray*}
ln \, det \, (\beta M) &=& ln\,det\, \beta \left[-G_0^{-1}(1-G_0 M_1)\right]\\
&=& ln\,det(-\beta G_0^{-1}) + Tr\,ln(1-G_0 M_1) \\
&=& ln\,det(-\beta G_0^{-1}) - 
Tr\{ \sum_{n=1}^{\infty} \frac{1}{n} (G_0 M_1)^n \}
\end{eqnarray*}
The first term $ln\,det(-\beta G_0^{-1})$ leads to the expression 
$\sum_i ln \, 2 ch \frac{\beta}{2} \| \vec{\varphi}_i(\omega=0) \|$. The first
term in the sum gives the contributions at the one-loop level.

\subsection{Relation between the Hubbard-Stratonovich mean fields 
$\bar{\varphi}_i$ and the mean-field magnetizations $\bar{m}_i$.
\label{ApB}}

Using  $\bar{m}_i = - \frac{\partial \mathcal{F}_{MF}}
{\partial B_i^z}$ one gets
\begin{eqnarray*}
\bar{\varphi}_j^z &=& \frac{2}{\beta} th^{-1} 2 \bar{m}_i \\
\bar{\varphi}_j^z - B_j &=& \sum_i J_{i,j}
\bar{m}_i \\
\frac{2}{\beta} th^{-1} 2 \bar{m}_i &=&
B_i + \sum_j J_{i,j} \bar{m}_j
\end{eqnarray*}

Starting from the N\'eel state  $\bar{m}_i = (-1)^i \bar{m} + (-1)^i 
\Delta \tilde{m}(B)$. For a weak magnetic field the second term can be 
linearized $\Delta \tilde{m}(B)=\Delta \tilde{m}_0 B$. The coupling strength 
matrix acting between nearest neighbour sites is taken as $J_{ij} = J 
\underset{\vec{\eta} \in n.n.}{\sum} \delta(\vec{r}_i-\vec{r}_j \pm \vec{\eta})$
with $J<0$. 
Then 

\begin{eqnarray*}
\bar{m}_i &=& (-1)^i \bar{m} + (-1)^i \Delta \tilde{m}(B) \\
&=& \frac{1}{2} th \frac{\beta}{2} \left[ (-1)^i B + Z |J| (-1)^i
(\bar{m}+\Delta \tilde{m}(B))\right] \\
(-1)^i (\bar{m}+\Delta \tilde{m}(B)) &=&
\frac{1}{2} th \frac{\beta}{2} (-1)^i \left[B + 2.D.|J|. 
(\bar{m}+\Delta \tilde{m}(B)) \right]
\end{eqnarray*}

By means of a Taylor expansion around $B=0$ :

\begin{eqnarray*}
(-1)^i(\bar{m}+\Delta \tilde{m}(B)) &=&
\frac{1}{2} th 
\frac{\beta}{2}(-1)^i.2.D|J|\bar{m}
 \\
&+& \frac{1}{2}.\frac{\beta}{2} (-1)^i
\left[1+2.D|J|\Delta\tilde{m}_0 \right]
\left[1-th^2 \frac{\beta}{2}(-1)^i 
2 D|J|\bar{m} \right].B
 \\
&+& \mathcal{O}(B^2)
\end{eqnarray*}
By identification one gets
$\Delta\tilde{m}_0 = \frac{\beta}{4} 
\left[1+2.D|J|\Delta\tilde{m}_0 \right]\left[1-4 \bar{m}^2 \right]$
and finally

\begin{gather*}
\bar{m} = \frac{1}{2} th \frac{\beta}{2}D|J|\bar{m} \\
\Delta \tilde{m}(B) = \Delta \tilde{m}_0.B \\
\Delta \tilde{m}_0 = \frac{\frac{\beta}{4}\left(1-4\bar{m}^2 \right)}
{1-\frac{\beta}{2}D|J|\left(1-4\bar{m}^2 \right)}
\end{gather*}
where $D$ is the lattice dimension.

\subsection{The free energy and the terms $\delta \varphi^z$ 
and $\delta \varphi^{+-}$ in equation \ref{delta2Seff} \label{ApC}}

Substituting $\bar{m}_i = (-1)^i(\bar{m}+\Delta \tilde{m}(B)$ in
$\frac{1}{2} \frac{\partial^2 S_{eff}}{{\partial \varphi^z}^2}
{\Big\vert}_{\left[\bar{\varphi}\right]}
 {\delta \varphi^z}^2$ and 
$\frac{1}{2} \frac{\partial^2 S_{eff}}{{\partial \varphi^{+}
\partial \varphi^{-}}}
{\Big\vert}_{\left[\bar{\varphi}\right]}
 {\delta \varphi^{+}} {\delta \varphi^{-}}$ 
of equation \ref{delta2Seff} leads to

\begin{eqnarray*}
(1) &=& \left(\frac{\beta}{4} 
th^{'} \left( \frac{\beta}{2} \bar{\varphi}_\alpha^z \right)\right)
  \\
&=& 
\frac{\beta}{4}\left(1- 4(\bar{m}+\Delta \tilde{m}(B))^2 \right)
\end{eqnarray*}

\begin{eqnarray*}
(2) &=& \left( \frac{1}{2} 
\frac{th \left( \frac{\beta}{2} \bar{\varphi}_\alpha^z \right) }
{\bar{\varphi}_\alpha^z - i \omega } \right)
 \\
&=& \left[ 2a \right]_\omega + (-1)^\alpha \left[ 2b \right]_\omega
 \\
\left[ 2a \right]_\omega &=&
\frac{(\bar{m}+\Delta \tilde{m}(B)).(B+2 D |J|(\bar{m}+\Delta \tilde{m}(B)))}
{\left[(B+2 D |J|(\bar{m}+\Delta \tilde{m}(B)))^2 + \omega^2 \right]}
 \\
\left[ 2b \right]_\omega &=&
\frac{i \omega (\bar{m}+\Delta \tilde{m}(B))}
{\left[(B+2 D |J|(\bar{m}+\Delta \tilde{m}(B)))^2 + \omega^2 \right]}
 \\
\end{eqnarray*}

Integrating out the fluctuations $\vec \delta \varphi$ away from the mean field 
$\bar{\varphi}^z$ leads to

\begin{gather*}
\bar{m} = \frac{1}{2} th \frac{\beta}{2}D|J|\bar{m}
 \\
\Delta \tilde{m}(B) = \Delta \tilde{m}_0.B
 \\
\Delta \tilde{m}_0 = \frac{\frac{\beta}{4}\left(1-4.\bar{m}^2 \right)}
{1-\frac{\beta}{2}D|J|\left(1-4.\bar{m}^2 \right)}
 \\
\begin{aligned}
\mathcal{F}_{MF} =& 
N_p D |J| \left( \bar{m}+\Delta \tilde{m}(B) \right)^2
- \frac{N_p}{\beta} \text{ln ch } \left(\frac{\beta}{2} 
\left[ B+2 D |J|(\bar{m}+\Delta \tilde{m}(B))\right] \right)
 \\
\delta\mathcal{F}_{zz} =&
\frac{1}{2\beta} \sum_{\vec{k} \in SBZ}
ln \left[ 1 - \left( \frac{\beta D |J| \gamma_{\vec{k}} }{2} \right)^2
\left[ 1- 4(\bar{m}+\Delta \tilde{m}(B))^2 \right]^2 \right]
 \\
\delta\mathcal{F}_{+-} =&
\frac{2}{\beta} \sum_{\vec{k} \in SBZ}
\text{ln} \Bigg{(}
\frac{\text{Sinh} \left(\frac{\beta}{2}
\left(\left[B + 2 D |J|(\bar{m}+\Delta \tilde{m}(B))\right]^2
- \left[2 D |J| \gamma_{\vec{k}} (\bar{m}+\Delta \tilde{m}(B))\right]^2 
\right)^{1/2} \right)}
{\text{Sinh} \left(\frac{\beta}{2} 
\left[ B+2 D |J|(\bar{m}+\Delta \tilde{m}(B))\right] \right)}
\Bigg{)}
\end{aligned}
\end{gather*}

\subsection{The XXZ-model \label{ApD}}

The XXZ Hamiltonian 
\begin{eqnarray*}
H^{XXZ} = -\frac{J}{2} \sum_{<ij>} \left( S_i^x S_j^x+ S_i^y S_j^y 
+ (1+\delta)S_i^z S_j^z \right)
\end{eqnarray*}
leads to a critical temperature

\begin{gather*}
T_c^{XXZ} = \frac{D |J+\Delta|}{2} \\
\Delta = J  \delta \\
\end{gather*}
and a mean magnetization
\begin{gather*}
\bar{m} = \frac{1}{2} th \frac{\beta}{2}D|J+\Delta|\bar{m} \\
\Delta \tilde{m}(B) = \Delta \tilde{m}_0.B
= B.\frac{\frac{\beta}{4}\left(1-4.\bar{m}^2 \right)}
{1-\frac{\beta}{2}D|J+\Delta|\left(1-4.\bar{m}^2 \right)}
\end{gather*}

\subsubsection{Free energy}
\begin{align*}
\mathcal{F}_{MF} =& 
N_p D |J+\Delta| \left( \bar{m}+\Delta \tilde{m}(B) \right)^2
- \frac{N_p}{\beta} \text{ln ch } \left(\frac{\beta}{2} 
\left[ B+2 D |J+\Delta|(\bar{m}+\Delta \tilde{m}(B))\right] \right)
 \\
\delta\mathcal{F}_{zz} =&
\frac{1}{2\beta} \sum_{\vec{k} \in SBZ}
ln \left[ 1 - \left( \frac{\beta D |J+\Delta| \gamma_{\vec{k}} }{2} \right)^2
\left[ 1- 4(\bar{m}+\Delta \tilde{m}(B))^2 \right]^2 \right]
 \\
\delta\mathcal{F}_{+-} =&
\frac{2}{\beta} \sum_{\vec{k} \in SBZ}
\text{ln} \Bigg{(}
\frac{\text{Sinh} \left(\frac{\beta}{2}
\left(\left[B + 2 D |J+\Delta|(\bar{m}+\Delta \tilde{m}(B))\right]^2
- \left[2 D |J| \gamma_{\vec{k}} (\bar{m}+\Delta \tilde{m}(B))\right]^2 
\right)^{1/2} \right)}
{\text{Sinh} \left(\frac{\beta}{2} 
\left[ B+2 D |J+\Delta|(\bar{m}+\Delta \tilde{m}(B))\right] \right)}
\Bigg{)}
\end{align*}

\subsubsection{Magnetization}
\begin{eqnarray*}
m &=& \bar{m} + \delta m_{zz} + \delta m_{+-}
 \\
\bar{m} &=& -\frac{1}{N_p} \frac{\partial \mathcal{F}_{MF}}{\partial B}
\Bigg{\vert}_{B=0}
 \\
\delta m_{zz} &=& -\frac{1}{N_p} \frac{\partial \mathcal{F}_{zz}}{\partial B}
\Bigg{\vert}_{B=0} \\
&=& -\frac{1}{N_p \beta} \sum_{\vec{k} \in SBZ}
\frac{8.\bar{m} \Delta \tilde{m}_0 \left(1-4\bar{m}^2\right)
\left(\frac{\beta D |J+\Delta| \gamma_{\vec{k}}}{2}\right)^2}
{\left[1 -\left(\frac{\beta D |J+\Delta| \gamma_{\vec{k}}}{2}\right)^2 
 \left(1-4\bar{m}^2\right)\right]}
 \\
\delta m_{+-} &=&
\frac{\left(1 + 2 D |J+\Delta| \Delta \tilde{m}_0 \right)}{4\bar{m}}
 \\
&\phantom{=}& 
- \frac{1}{N_p} \sum_{\vec{k} \in SBZ}
\frac{\left(1 + 2 D |J+\Delta| \Delta \tilde{m}_0 
(1 -\left(\frac{J}{J+\Delta} \gamma_{\vec{k}}\right)^2)\right)}
{\sqrt{1 - \left(\frac{J}{J+\Delta} \gamma_{\vec{k}}\right)^2}}
\frac{1}{\left[\text{th} \left(\beta D |J| \bar{m}
\sqrt{1 - \left(\frac{J}{J+\Delta} \gamma_{\vec{k}}\right)^2} \right) \right]}
\end{eqnarray*}

\subsubsection{Susceptibility}

\begin{align*}
\chi_\parallel 
=& -\frac{1}{N_p} \frac{\partial^2 \mathcal{F}}{\partial B^2} 
\Bigg{\vert}_{B=0} =\chi_{MF} + \chi_{zz} + \chi_{+-}
 \\
{\chi_\parallel}_{MF} 
=& \Delta m_{\chi 0} = \frac{ \frac{\beta}{4}\left( 1-4.\bar{m}^2 
\right) }{1+\frac{\beta}{2}D|J+\Delta|\left(1-4\bar{m}^2 \right)}
 \\
{\chi_\parallel}_{zz}
=& -\frac{1}{N_p \beta} \sum_{\vec{k} \in SBZ}
\frac{8\left( \frac{\beta D |J+\Delta| \gamma_{\vec{k}} }{2} \right)^2
\Delta m_{\chi 0}^2 \left(1 + 4 \bar{m}^2 \right)  }
{\left[ 1 - \left( \frac{\beta D |J+\Delta| \gamma_{\vec{k}} }{2} \right)^2 
\left( 1 - 4 \bar{m}^2 \right)^2 \right]}
 \\
{\chi_\parallel}_{+-}
=&
 \frac{1}{N_p} \sum_{\vec{k} \in SBZ} \Bigg{\{}
- \frac{1}{2} \frac{\beta \left(1 - 2 D |J+\Delta| \Delta m_{\chi 0} \right)^2}
{\text{sh }^2 \left(\beta D |J+\Delta| \bar{m} \right)}
 \\
& + \frac{1}{\text{sh }^2 \left(\beta D |J+\Delta| 
\bar{m}\sqrt{1-\left(\frac{J}{J+\Delta}\gamma_{\vec{k}}\right)^2} \right)} 
\Bigg{[}
\frac{\beta}{2} \left(1 - 2 D |J+\Delta| \Delta m_{\chi 0} \right)^2
 \\
& - \beta \left(D |J| \Delta m_{\chi 0} \gamma_{\vec{k}} \right)^2 
\frac{
\text{sh }2\beta D |J+\Delta| \bar{m}
\sqrt{1-\left(\frac{J}{J+\Delta}\gamma_{\vec{k}}\right)^2}
 }
{
\beta D |J+\Delta| \bar{m}
\sqrt{1-\left(\frac{J}{J+\Delta}\gamma_{\vec{k}}\right)^2} }
\Bigg{]}
\Bigg{\}}
\end{align*}

\end{document}